\documentclass{ws-procs975x65}

\begin{document}

\markboth{Aar\'{o}n V. B. Arellano and Francisco S. N.
Lobo}{Dynamic wormhole spacetimes coupled to nonlinear
electrodynamics}

\wstoc{Dynamic wormhole spacetimes coupled to nonlinear
electrodynamics}{Aar\'{o}n V. B. Arellano and Francisco S. N.
Lobo}

\title{DYNAMIC WORMHOLE SPACETIMES COUPLED TO NONLINEAR ELECTRODYNAMICS}

\author{AAR\'{O}N V. B. ARELLANO}
\address{Facultad de Ciencias,
Universidad Aut\'{o}noma del Estado de M\'{e}xico, \\
El Cerrillo, Piedras Blancas, C.P. 50200, Toluca, M\'{e}xico\\
\email{vynzds@yahoo.com.mx} }

\author{FRANCISCO S. N. LOBO}
\address{Centro de Astronomia
e Astrof\'{\i}sica da Universidade de Lisboa,\\
Campo Grande, Ed. C8 1749-016 Lisboa, Portugal\\
\email{flobo@cosmo.fis.fc.ul.pt}}

\begin{abstract}

We explore the possibility of dynamic wormhole geometries, within
the context of nonlinear electrodynamics. The Einstein field
equation imposes a contracting wormhole solution and the obedience
of the weak energy condition. Furthermore, in the presence of an
electric field, the latter presents a singularity at the throat,
however, for a pure magnetic field the solution is regular. Thus,
taking into account the principle of finiteness, that a
satisfactory theory should avoid physical quantities becoming
infinite, one may rule out evolving wormhole solutions, in the
presence of an electric field, coupled to nonlinear
electrodynamics.

\end{abstract}

\keywords{Traversable wormholes; nonlinear electrodynamics.}

\bodymatter

%-----------------------------------------------
\section*{}
%-----------------------------------------------

Pioneering work on nonlinear electrodynamic theories may be traced
back to Born and Infeld \cite{BI}, where the latter outlined a
model to remedy the fact that the standard picture of a point
charged particle possesses an infinite self-energy. Therefore, the
Born-Infeld model was founded on a principle of finiteness, that a
satisfactory theory should avoid physical quantities becoming
infinite. Recently, nonlinear electrodynamics has found a wide
range of applicability, namely, as effective theories at different
levels of string/M-theory, cosmological models, black holes, and
in wormhole physics, amongst others (see
\cite{Arellano1,Arellano2} and references therein).

Relatively to wormhole physics it was found that static
spherically symmetric and stationary axisymmetric traversable
wormholes cannot exist within nonlinear electrodynamic, mainly due
to the presence of event horizons, the non-violation of the null
energy condition at the throat, and due to the imposition of the
principle of finiteness \cite{Arellano2,Bronnikov1}. In this work,
we shall explore the possibility that nonlinear electrodynamics
may support time-dependent traversable wormhole geometries. This
is of particular interest as the energy conditions are not
necessarily violated for evolving wormhole spacetimes \cite{Kar1}.

The action of $(3+1)-$dimensional general relativity coupled to
nonlinear electrodynamics is given by (with $G=c=1$)
\begin{equation}
S=\int \sqrt{-g}\left[\frac{R}{16\pi}+L(F)\right]\,d^4x  \,,
\end{equation}
where $R$ is the Ricci scalar. $L(F)$ is a gauge-invariant
electromagnetic Lagrangian, depending on a single invariant $F$
given by $F=F^{\mu\nu}F_{\mu\nu}/4$, where $F_{\mu\nu}$ is the
electromagnetic tensor. Note that in Einstein-Maxwell theory, the
Lagrangian takes the form $L(F)= -F/4\pi$.

Varying the action with respect to the gravitational field
provides the Einstein field equations $G_{\mu\nu}=8\pi
T_{\mu\nu}$, with the stress-energy tensor given by
\begin{equation}
T_{\mu\nu}=g_{\mu\nu}\,L(F)-F_{\mu\alpha}F_{\nu}{}^{\alpha}\,L_{F}\,,
    \label{4dim-stress-energy}
\end{equation}
where $L_F=dL/dF$.

We shall consider that the spacetime metric representing a dynamic
spherically symmetric $(3+1)-$dimensional wormhole, which is
conformally related to the static wormhole geometry \cite{Morris},
takes the form
\begin{equation} \label{4dysme}
ds^2=\Omega^2(t)\left[-e ^{2\Phi(r)}dt^2+\frac{dr^2}{1-
b(r)/r}+r^2(d\theta ^2+\sin ^2{\theta}d\phi ^2)\right]  \,,
\end{equation}
where $\Phi$ and $b$ are functions of $r$, and $\Omega=\Omega(t)$
is the conformal factor, which is finite and positive definite
throughout the domain of $t$. To be a wormhole solution, the
following conditions are imposed: $\Phi(r)$ is finite everywhere
in order to avoid the presence of event horizons; $b(r)/r<1$, with
$b(r_0)=r_0$ at the throat; and the flaring out condition
$(b-b'r)/b^2 \geq 0$, with $b'(r_0)<1$ at the throat.

For this particular case, the weak energy condition, which is
defined as $T_{\mu\nu}U^{\mu}U^{\nu}\geq0$ where $U^{\mu}$ is a
timelike vector, is satisfied \cite{Arellano1}, contrary to the
static and spherically symmetric traversable wormholes
\cite{Arellano2,Bronnikov1}.

Through the Einstein field equation, we obtain the following
relationship
\begin{equation}\label{4boeq}
\frac{b'r-b}{2r^3}=-\left[2(\dot\Omega/\Omega)^2
-\ddot\Omega/\Omega\right]
  \,,
\end{equation}
which provides the solutions
\begin{equation}
  b(r)=r\left[1-\alpha^2(r^2-r_0^2)\right]    \,,
\qquad
  \Omega(t)=\frac{2\alpha}{C_1\,e^{\alpha t}-C_2\,e^{-\alpha t}}
  \,,\label{4solom}
\end{equation}
where $\alpha$ is a constant, and $C_1$ and $C_2$ are constants of
integration. Now, $\Omega(t) \rightarrow 0$ as $t\rightarrow
\infty$, which reflects a contracting wormhole solution. This
analysis shows that one may, in principle, obtain an evolving
wormhole solution in the range of the time coordinate.
A fundamental condition to be a solution of a wormhole, is that
$b(r)>0$ is imposed\cite{Lemos}. Thus, the range of $r$ is
$r_0<r<a=r_0\sqrt{1+1/\beta^2}$, with $\beta=\alpha r_0$. If $a
\gg r_0$, i.e., $\beta\simeq r_0/a\ll 1$, one may have an
arbitrarily large wormhole. Note, however, that one may, in
principle, match this solution to an exterior vacuum solution at a
junction interface $R$, within the range $r_0<r<a$.

The electromagnetic field equations take the following form
\begin{equation}
\left(F^{\mu\nu}\,L_{F}\right)_{;\mu}=0 \,, \qquad
\left(^*F^{\mu\nu}\right)_{;\mu}=0   \,.
     \label{em-field}
\end{equation}
Taking into account the symmetries of the geometry, the non-zero
compatible terms for the electromagnetic tensor are
$F_{\mu\nu}=2E(x^\alpha)\,\delta^{[t}_\mu
\,\delta^{r]}_\nu+2B(x^\alpha)\,\delta^{[\theta}_\mu
\,\delta^{\phi]}_\nu \label{em-tensor}$, where $F_{tr}=E$ is the
electric field, and $F_{\theta\phi}=B$, the magnetic field. From
the electromagnetic field equations, we deduce the following
\begin{equation}\label{4NE}
  E(t,r)=\frac{f\,\Omega^2r\pm\sqrt{f^2\Omega^4r^2
  -(32\pi q_{\rm e} q_{\rm m})^2}}{32\pi q_{\rm e}
  r^2(1-b/r)^{1/2}}\,, \qquad   B(\theta)=q_{\rm m}\,\sin\theta
  \,,
\end{equation}
with $f=(b'r-3b)$, and $q_{\rm e}$ and $q_{\rm m}$ are constants
related to the electric and magnetic charge, respectively. From
this solution we point out two observations: (i) the requirement
of $f^2\Omega^2r^2>(32\pi q_{\rm e}q_{\rm m})^2$; (ii) and
$E\propto (1-b/r)^{-1/2}$, showing that the $E$ field is singular
at the throat, which is in contrast to the principle of
finiteness.

An interesting case arises considering a pure magnetic field,
$E=0$, from which we obtain the Lagrangian and its derivative
\begin{equation}\label{4NBLF}
L=-\frac{1}{8\pi\Omega^2}\left[b'/r^2
+3(\dot{\Omega}/\Omega)^2\right] \,, \qquad
  L_{F}=\frac{1}{16\pi q_{\rm m}^2}\Omega^2r(b'r-3b)\,.
\end{equation}
These equations, together with $B=q_{\rm m}\sin\theta$, $F=q_{\rm
m}^2/(2\Omega^4r^4)$ and solutions (\ref{4solom}) provide a
regular wormhole solution at the throat, with finite fields. We
emphasize that this result is in close relationship to the regular
magnetic black holes coupled to nonlinear electrodynamic found by
Bronnikov \cite{Bronnikov1}.

In conclusion, we have explored the possibility of evolving
time-dependent wormhole geometries coupled to nonlinear
electrodynamics. It was found that the Einstein field equation
imposes a contracting wormhole solution and that the weak energy
condition is satisfied. In the presence of an electric field, a
problematic issue was verified, namely, that the latter becomes
singular at the throat. However, regular solutions of traversable
wormholes in the presence of a pure magnetic field were found.
Another point worth noting is that we have only considered that
the gauge-invariant electromagnetic Lagrangian $L(F)$ be dependent
on a single invariant $F$. It would also be worthwhile to include
another electromagnetic field invariant $G\sim
\,^*F^{\mu\nu}\,F_{\mu\nu}$, which would possibly add an
interesting analysis to the solutions found in this work.

%------------------------------------

%-----------------------------------------------------------
% References
%-----------------------------------------------------------

\vfill
%\pagebreak

\end{document}